\documentstyle[12pt]{article}
\newlength{\bredde}
\def\slash#1{\settowidth{\bredde}{$#1$}\ifmmode\,\raisebox{.15ex}{/}
\hspace*{-\bredde} #1\else$\,\raisebox{.15ex}{/}\hspace*{-\bredde} #1$\fi}
\newcommand{\beq}{\begin{equation}}
\newcommand{\eeq}{\end{equation}}
\newcommand{\wt}{\widetilde}

\def\gtwid{\raise.3ex\hbox{$>$\kern-.75em\lower1ex\hbox{$\sim$}}}
\def\ltwid{\raise.3ex\hbox{$<$\kern-.75em\lower1ex\hbox{$\sim$}}}
\begin{document}
\vspace*{1cm}
\topmargin -0.8cm
\oddsidemargin -0.8cm
\evensidemargin -0.8cm
\headheight 0pt
\headsep 0pt
\topskip 9mm
\title{\Large{
Schwinger-Dyson BRST-Symmetry and the Equivalence of Hamiltonian and
Lagrangian Quantisation}}
\vspace{0.5cm}

\author{{\sc Frank De Jonghe}\thanks{
Aspirant of the N.F.W.O., Belgium.}\\
Instituut voor Theoretische Fysica, K.U.Leuven, Belgium \\
  and \\
CERN -- Geneva
}
\maketitle
\vfill
\begin{abstract} Implementing the requirement that a field
theory be invariant under Schwinger-Dyson BRST symmetry in the
Hamiltonian
formalism, we show the equivalence between
Hamiltonian and Lagrangian BRST-formalism at the path integral level.
The Lagrangian
quantum master equation is derived as a direct consequence of the
Fradkin-Vilkovisky
theorem in Hamiltonian BRST quantisation.
\end{abstract}
\vspace{8.5cm}
\begin{flushleft}
KUL-TF-93/08 \\
CERN--TH-6823/93 \\
March 1993
\end{flushleft}
\thispagestyle{empty}
\newpage

There are two approaches available for the quantisation of
gauge theories, both
rooted in the concept of BRST symmetry. First, a Hamiltonian method was
developed (below denoted by BFV)
 \cite{BFV}, with as key features a dynamical treatment of the Lagrange
multipliers
imposing the constraints, the introduction of a canonically conjugate
pair of a ghost and an antighost for every first-class constraint,
the existence of a nilpotent fermionic BRST-generator, and the existence
of a BRST invariant Hamiltonian. Using these ingredients one can
construct a path integral which is independent of the gauge fixing,
as guaranteed by the celebrated Fradkin-Vilkovisky theorem \cite{BFV}.
Guided by but
 not as a logical consequence of this Hamiltonian formalism a Lagrangian
formalism
was developed (below denoted by BV)\cite{BV}.
Here, the most important property is the doubling of the fields,
by associating
with every field $\phi^a$
an antifield $\phi^*_a$ of opposite statistics. Solving the
so-called master equation gives an extended action, depending on
both the fields and the anti-fields. The gauge fixing is done by
replacing the antifield $\phi^*_a$ by $\frac{\delta \Psi(\phi)}{\delta
\phi^a}$, where $\Psi$ is a fermionic function of the fields, called
the gauge fermion \cite{BFV, BV}.
In this paper we will
prove the equivalence of both methods for quantisation.

A large effort has already been devoted over the years to precisely this
problem.
In a series of papers \cite{Bruss}, Henneaux et al. proved the following.
Given any Lagrangian, one can try to find its gauge symmetries directly,
and
treat them in the scheme of BV. On the other hand, one could look for the
gauge
symmetries using the Hamiltonian methods of Dirac. The canonical action so
obtained
can again be treated in the BV scheme. Both methods were proven to lead
to the same path integral. The most remarkable feature of this approach is
that in the Hamiltonian method, anti-fields have to be introduced for the
the momenta. The Lagrangian analog of the Hamiltonian BRST-charge has also
been constructed in \cite{Barca}.

The procedure we will follow here, is closer to that of
\cite{GriGriT}. There it was
shown that starting from a complete Hamiltonian analysis following BFV, it
is
possible to obtain a Lagrangian action satisfying the quantum master
equation
of BV, after integrating out the momenta. However, the anti-fields have to
be
introduced in the Hamiltonian path integral in a rather ad hoc way as
sources
for the BRST-transformation. Also, the Lagrangian master equation requires
a seperate proof, where one would expect it to follow naturally from the
Fradkin-Vilkovisky theorem, as they both have as consequence that the
generating function does not depend on the choice
gauge. We will clarify these two points below.

Recently, it was shown \cite{us} that demanding a Lagrangian
field theory to be invariant under Schwinger-Dyson BRST symmetry \cite
{us0}
in addition to usual BRST symmetries of internal gauge invariances
is a natural way of introducing the anti-fields of
the Batalin-Vilkovisky Lagrangian formalism. It turns out that they are
the anti-ghosts of the Schwinger-Dyson BRST symmetry.
The derivation is done most simply in terms of
a certain Lagrangian collective field formalism, described in
detail in ref. \cite{us0}, but can also be based directly on the
Schwinger-Dyson BRST symmetry itself. However, it is less obvious how
to implement the latter approach in the Hamiltonian formalism.

Our strategy to prove the equivalence of the Hamiltonian BFV formalism
and the Lagrangian BV formalism will then be to
consider a Hamiltonian system
with known
extended phase space, BRST-charge and BRST-invariant Hamiltonian.
These are all
constructed following the prescriptions of the BFV formalism
\cite{GovHen}.
Then we implement the Schwinger-Dyson
shift symmetry in a way consistent with the original symmetries.
Extending further the extended phase space to include the ghost-antighost
pairs for this shift symmetry, we introduce the anti-fields.
Gauge fixing the shift-symmetry and the original gauge-symmetries allows
one to construct the path-integral. After integrating out all momenta,
except the ones for the Lagrange multipliers of the Hamiltonian system,
one ends up with an action satisfying the BV quantum master equation.
Anti-fields indeed act as sources for BRST-transformations and the
left-over momenta become the Nakanishi-Lautrup fields.

We will first recapitulate how Schwinger-Dyson shift symmetries are to be
described in the Hamiltonian formalism \cite{us0}.
To that purpose we first consider a
Lagrangian depending on fields and their time-derivatives -denoted by a
dot- which has no internal gauge symmetries :
$L(t) = L (\phi^a, \dot{\phi}^a) $. Introducing collective fields
amounts here to considering $ L(\phi^a- \varphi^a, \dot{\phi}
^a - \dot{\varphi}^a ) $.
Denoting the momenta conjugate to $\phi^a$ by $\pi_a$ and to $\varphi^a$
by $\varpi_a$, the first class constraints are
\beq \chi_a = \pi_a + \varpi_a = 0 \eeq
{}From this, it trivially follows that the structure constants $C^a_{bc}$
vanish,
 $[ \chi_b , \chi_c ] = 0$, as $ [ \pi_a , \phi^b ] = [\varpi_a , \varphi
^b ] = - \delta_a^b $ are the only non-vanishing Poisson brackets.
These constraints also have a vanishing Poisson
bracket with the Hamiltonian, as the latter also only depends on the
difference
$\phi^a - \varphi^a$. This is equivalent to saying that the structure
constants $V^b_a$ vanish.
After associating one Lagrange multipier $\lambda^a$
and its canonical momentum $\pi_{\lambda a}$
 ( $[ \pi_{\lambda a} , \lambda^b ] = - \delta_a^b $ ) with every first
class constraint $\chi_a$
the complete set of constraints $G_a = ( \pi_{\lambda a} , \chi_a )$ still
has vanishing structure constants. We now introduce the following ghost
and antighost fields to construct the extended phase space \cite{GovHen}
\begin{eqnarray}
     \pi_{\lambda a} = 0  & \stackrel{gh\sharp=\:1}{\leftrightarrow} &
      -i^{a+1} {\cal P}^a  \nonumber \\
 & \stackrel{gh\sharp=\:-1}{\leftrightarrow}
                & i^{a+1} \phi^*_a  \nonumber \\
     \chi_a = 0 & \stackrel{gh\sharp=\:1}{\leftrightarrow}
   &  c^a \nonumber \\
 & \stackrel{gh\sharp=\:-1}{\leftrightarrow}
              &  \bar{{\cal P}}_a          \label{extps}
\end{eqnarray}
with the
only non-vanishing brackets $ [\bar{{\cal P}}_a,c^b ] = [ {\cal P}^b ,
\phi^*_a ] = - \delta^b_a $. The ghostfields (those of $gh\sharp=1$)
are collectively denoted
by $\eta^a$, the anti ghosts ($gh\sharp=-1$) by $\theta_a$, including
the numerical factors $i$ in (\ref{extps}).
Using this, one can straightforwardly construct the BRST generator of the
shift symmetries
\beq
\Omega_s = -i^{a+1}{\cal P}^a \pi_{\lambda a} + c^a ( \pi_a + \varpi_a)
\eeq
Gauge fixing the collective field to zero can then be done by taking as
gauge-fermion \beq
 \psi = \bar{{\cal P}}_a \lambda^a + \frac{i^{a+1}}{\beta} \phi^*_a
\varphi^a
\label{phizero} \eeq
where $\beta$ is an arbitrary parameter.
We will take the $\beta \rightarrow
0$ limit later.
The Poisson bracket with the BRST-charge then gives
\beq [ \psi , \Omega_s ] = i^{a+1} {\cal P}^a \bar{{\cal P}}_a
        - \frac{1}{\beta} \varphi^a \pi_{\lambda a} - \lambda^a (\pi_a +
\varpi_a ) + \frac{i^{a+1} (-1)^a}{\beta} \phi^*_a c^a \eeq
The action to be exponentiated in the path-integral is then according
to the standard BFV formalism given by
\beq S = \dot{\phi}^a \pi_a + \dot{\varphi}^a \varpi_a
+ \dot{\lambda}^a \pi_{\lambda a}  + \dot{\eta}^a \theta_a
 - H ( \pi , \varpi , \phi - \varphi ) + [ \psi , \Omega_s ]
\label{actie} \eeq
We now redefine
\begin{eqnarray}
      \frac{1}{\beta} \pi_{\lambda a} & \rightarrow & \pi_{\lambda a}
      \nonumber \\
      \frac{i^{a+1}}{\beta} \phi^*_a & \rightarrow & \phi^*_a
        \nonumber \\
      (-1)^a c^a & \rightarrow & c^a
   \label{redef}
\end{eqnarray}
At this moment we take the limit $\beta \rightarrow 0$, which makes
two terms of the symplectic part of S disappear. After that, the integral
over the ghostmomenta
 ${\cal P}^a$ and $\bar{{\cal P}}_a$ becomes trivial, leading to
\beq
 S = \dot{\phi}^a \pi_a + \dot{\varphi}^a \varpi_a - H (\pi , \varpi ,
 \phi
-\varphi) - \varphi^a \pi_{\lambda a}  - \lambda^a ( \pi_a + \varpi_a )
 + \phi^*_a c^a \eeq
By integrating over the Lagrange multipliers and their momenta, the
path integral becomes
\beq
\int [d\phi^a] [d c^a ] [d \phi^*_a]  e^{\frac{i}{\hbar} \left[ {\cal S}
  + \phi^*_a  c^a \right] } \eeq
where
\beq
e^{\frac{i}{\hbar} {\cal S}}= \int [d \pi_a ] \exp \left[\frac{i}{\hbar}
\left( \int dt \dot{\phi}^a \pi_a -
H (\pi , -\varpi , \phi ) \right) \right] \eeq
This parallels the result of \cite{us} for the case of no gauge-symmetries
{}.
The action is nothing but the original action, plus a term
where the anti-fields multiply the
ghosts $c^{a}$ of Alfaro and Damgaard \cite{us}.
Notice also that the collective field has been eliminated completely.
Recently, it was shown \cite{me} that this is not possible when one
tries to construct a path integral which is invariant under both
BRST and anti-BRST transformations. The collective field then is needed
as a source for mixed transformations \cite{BLT}.

We want to add another comment, which has some bearing on the case
where there are originally gauge-symmetries present. The Hamiltonian
introduced in (\ref{actie}) has a rather arbitrary dependence on the
momenta. For instance, if we had started from a Lagrangian quadratic
in the velocities, $L(t) = \frac{1}{2} \dot{\phi}^2 + \wt{L}(\phi)$ then
an arbitrary parameter could be present in the Hamiltonian
\beq
H = \frac{1}{2} \mu \pi^2 +\frac{1}{2} (1-\mu) \varpi^2 - \wt{L}(\phi
- \varphi)            \label{Hmu}               \eeq
Of course, the $\mu$-dependence disappears when imposing the constraint.
When there are gauge symmetries present, a specific choice of
$\mu $ is needed to make contact with BV.

Now we turn to the case of a Hamiltonian with gauge symmetries. We start
from
an extended phase space, whose coordinates ($\phi^{\alpha}, c^a,
 \bar{c}_a, \lambda^a$
for the case of first class irreducible theories) we collectively denote
by
$Y^{\bar{a}}$, the conjugate momenta $(\pi_{\alpha}, \bar{{\cal P}}_a,
{\cal P}^a, \pi_{\lambda a})$ by $\Pi_{\bar{a}}$.
The fundamental Poisson bracket is $[\Pi_{\bar{a}}, Y^{\bar{b}} ] =
-\delta
_{\bar{a}}^{\bar{b}}$. The fact that the symplectic form might be more
complicated after the elimination of second-class constraints, can easily
be incorporated.
We start from a known BRST-charge $\Omega_0(\Pi_{\bar{a}}, Y^{\bar{a}})$
and BRST-invariant Hamiltonian $H_0(\Pi_{\bar{a}}, Y^{\bar{a}})$
that satisfy the standard conditions  $ [ \Omega_0 , \Omega_0 ] =
[ H_0 , \Omega_0 ] = 0 $.

We now introduce for every field $Y^{\bar{a}}$
a collective field ${\cal Y}^
{\bar{a}}$ which has conjugate momenta $\Upsilon_{\bar{a}}$, again with
fundamental Poisson bracket $ [ \Upsilon_{\bar{a}} , {\cal Y}^{\bar{b}}] =
- \delta_{\bar{a}}^{\bar{b}}$. It will be useful to associate with every
phase space function $F (\Pi_{\bar{a}},Y^{\bar{a}}) $ a function $ \wt{F}
= F (- \Upsilon_{\bar{a}} , Y^{\bar{a}} - {\cal Y}^{\bar{a}})$. It is then
straightforward to show that \beq [ \wt{F} , \wt{G} ] = \wt{[F,G]}
\label{FtGt=FGt} \eeq
As the constraints of the shift symmetry are still given by
$ \chi_{\bar{a}} = \Pi_{\bar{a}} + \Upsilon_{\bar{a}} $, it also holds
that \beq [ \wt{F} , \chi_{\bar{a}} ] = 0  \label{FtC} \eeq

In order to construct the correct Hamiltonian
for which both the original gauge
symmetries and the shift symmetries are present, we look for
\begin{eqnarray}
        \Omega = \wt{\Omega_0} + \Omega_s \\
        H = \wt{H_0} + \Delta H
\end{eqnarray}
and demand that $[ \Omega , \Omega ] = [H ,\Omega ] = 0$.

Constructing the extended phase space as in the case of no internal
symmetries
\begin{eqnarray}
  \pi_{\lambda \bar{a}} = 0  & \stackrel{gh\sharp=\:1}{\leftrightarrow} &
      -i^{\bar{a}+1} {\cal P}^{\bar{a}} \nonumber \\
 & \stackrel{gh\sharp=\:-1}{\leftrightarrow}
                & i^{\bar{a}+1} Y^*_{\bar{a}} \nonumber \\
     \chi_a = 0 & \stackrel{gh\sharp=\:1}{\leftrightarrow}
    & c^{\bar{a}}  \nonumber \\
 & \stackrel{gh\sharp=\:-1}{\leftrightarrow}
              &  \bar{{\cal P}}_{\bar{a}}
\end{eqnarray}
the total BRST-charge is obtained by taking for $\Omega_s$ the same thing
as in the case without gauge symmetries.
\beq
 \Omega = \wt{\Omega_0} -i^{\bar{a}+1}{\cal P}^{\bar{a}} \pi_{\lambda
\bar{a}}
 + c^{\bar{a}} ( \Pi_{\bar{a}} + \Upsilon_{\bar{a}}) \eeq
That this quantity satisfies $ [ \Omega , \Omega ] = 0 $
follows immediately
from (\ref{FtGt=FGt}) and (\ref{FtC}).
{}From these it is also
clear that $ \Delta H = 0 $, i.e. $H = \wt{H_0}$. In fact, these two
properties
simply reflect the vanishing of the structure constants associated with
the Poisson
brackets of the original constraints and the original Hamiltonian with the
constraints of the shift symmetry, when the former are evaluated in
$(-\Upsilon_{\bar{a}} , Y^{\bar{a}} - {\cal Y}^{\bar{a}})$.

One might have wondered why the $\tilde{F}$ operation also involves a
substitution
of the momenta by minus the momenta of the collective field. This becomes
clear
when evaluating how the fields and the collective fields transform under
the
new BRST-charge.
The fields only transform under the shift symmetry
\beq [ Y^{\bar{a}} , \Omega ]  =  (-1)^{\bar{a}} c^{\bar{a}} \eeq
while the original gauge transformations have shifted to the collective
fields \beq
[ {\cal{Y}}^{\bar{a}},\Omega ] =  (-1)^{\bar{a}} c^{\bar{a}} + [ {\cal Y}
 ^{\bar{a}} , \wt{\Omega_0} ] \eeq
The original transformation can be shuffled between the fields and their
collective partners, which is related to the arbitrariness in the
momentum dependence of the Hamiltonian, as exemplified in (\ref{Hmu}).
It is precisely by our momentum substitution
rule that they end up entirely in the collective field transformation.
It was shown in \cite{us} that this corresponds to the BV boundary
condition that the solution of the classical master equation starts with
terms linear in the anti-fields, i.e. source terms for the BRST
formations. Other choices lead to equally well-defined path integrals
but they do not necessarilly agree with the BV boundary conditions.

For the gauge-fixing of the collective field we use again (\ref{phizero})
\beq
 \psi = \bar{{\cal P}}_{\bar{a}} \lambda^{\bar{a}} + \frac{i^{\bar{a}+1}}
    {\beta}  Y^*_{\bar{a}} {\cal Y}^{\bar{a}}  \eeq
leading to \beq
  [ \psi , \Omega ] = i^{\bar{a}+1} {\cal P}^{\bar{a}}
\bar{{\cal P}}_{\bar
{a}} - \frac{1}{\beta} {\cal Y}^{\bar{a}} \pi_{\lambda \bar{a}}
- \lambda^{\bar{a}} ( \Pi_{\bar{a}} + \Upsilon_{\bar{a}} )
+\frac{i^{\bar{a}+1}
(-1)^{\bar{a}}}{\beta} Y^*_{\bar{a}} c^{\bar{a}} +
 + \frac{i^{\bar{a}+1}}{\beta}
Y^*_{\bar{a}} [ {\cal Y}^{\bar{a}} , \wt{\Omega_0} ] \eeq
where the last term is the only, but important, difference from the case
of no gauge symmetries. We can rewrite its Poisson bracket as
\beq [ {\cal Y}^{\bar{a}} , \wt{\Omega_0} ] =  [ \wt{Y^{\bar{a}}} , \wt
{\Omega_0} ] = \wt{ [ Y^{\bar{a}} , \Omega_0 ] } \eeq
because $\wt{\Omega_0}$ does not depend on the momentum conjugate to $Y^
{\bar{a}} $.

The gauge fixing of the original symmetry can be done by taking a fermion
$ \Psi(\Pi_{\bar{a}} , Y^{\bar{a}})$ and adding
\beq [ \wt{\Psi} , \Omega ] = \wt{ [ \Psi , \Omega_0 ] } \label{gf}\eeq
Doing the same field-redefinitions as in (\ref{redef}), of course with
a replacement of the indices $a$ by $\bar{a}$, we get the
standard BFV action
\begin{eqnarray}
   S & = & \dot{Y}^{\bar{a}} \Pi_{\bar{a}} + \dot{{\cal Y}}^{\bar{a}}
   \Upsilon_
     {\bar{a}} + \beta \dot{\lambda}^{\bar{a}} \pi_{\lambda \bar{a}} -
     \beta i^{\bar{a}+1}
      \dot{{\cal P}}^{\bar{a}} Y^*_{\bar{a}} + (-1)^{\bar{a}} \dot{c}^
     {\bar{a}} \bar{{\cal P}}_{\bar{a}} \\
    &  & - \wt{H_0} + i^{\bar{a}+1} {\cal P}^{\bar{a}} \bar{{\cal P}}_
    {\bar{a}} - {\cal Y}^{\bar{a}} \pi_{\lambda \bar{a}}
     - \lambda^{\bar{a}} ( \Pi_{\bar{a}} + \Upsilon_{\bar{a}} )
      \nonumber \\
    &  & + Y^*_{\bar{a}} c^{\bar{a}} + Y^*_{\bar{a}} \wt
    { [ Y^{\bar{a}} , \Omega_0 ] } + \wt {[ \Psi , \Omega_0 ] }
        \nonumber
\end{eqnarray}
After taking the limit $\beta \rightarrow 0 $,
the ghost-momenta ${\cal P}$ and
$\bar{\cal P}$ can be integrated out trivially.
Doing then the integral over
$ \pi_{\lambda \bar{a}}$ gives
 a delta-function $\delta({\cal Y}^{\bar{a}})$,
while integrating out the Lagrange multiplier $\lambda^{\bar{a}}$ itself
leads
to $\delta(\Pi_{\bar{a}}+\Upsilon_{\bar{a}})$. It is trivial to see
that
imposing these delta-function constraints allows us to drop the tildes, so
that after integrating over the collective field and its momentum
we get \beq
 S = \dot{Y}^{\bar{a}} \Pi_{\bar{a}} - H_0 + Y^*_{\bar{a}}
  [ Y^{\bar{a}} , \Omega_0 ] + Y^*_{\bar{a}} c^{\bar{a}}
 + [\Psi , \Omega_0  ]
\eeq

We shall now show that the path integral obtained from this action, is
equal to the one obtained by doing Lagrangian BV analysis.
We make the most popular choice for the gauge fermion
\footnote{The Lagrange multipliers, ghosts
and anti ghosts that appear below, must not be confused
with the same quantities for the shift
symmetries which have been integrated over above. The former have indices
$a$, while the latter had ${\bar{a}}$. The point is that to proceed
further we have for the first time to specify in detail what fields
are contained in $Y^{\bar{a}}$ and $\Pi_{\bar{a}}$.}
\beq \Psi = \Psi_0( Y^{\bar{a}} , \pi_{\lambda a} ) + \bar{{\cal P}}_a
\lambda^a
 \eeq
Furthermore, we know the form of $\Omega_0$
\beq \Omega_0 = -i^{a+1} {\cal P}^a \pi_{\lambda a} + \Omega_{min} \eeq
where $\Omega_{min}$ does not depend on the Lagrange multiplier
$\lambda^a$ nor its momentum $\pi_{\lambda a}$. This allows us to write
\beq [\Psi , \Omega_0 ] = \frac{\delta^r \Psi_0}{\delta Y^{\bar{a}}}
[ Y^{\bar{a}}, \Omega_0 ] + i^{a+1} {\cal P}^a \bar{{\cal P}}_a
- (-1)^a  \frac{\delta^r \Omega_{min}}{\delta c^a} \lambda^a \eeq

It is now convenient to define $\hat{S}$ by
\beq S =  \hat{S} (Y^{\bar{a}} , Y^*_{\bar{a}}, \pi_{\lambda a} ,
 {\cal P}^a , \bar{{\cal P}}_a , \pi_{\alpha} )
+ Y^*_{\bar{a}} c^{\bar{a}} + \frac{\delta^r \Psi_0}
{\delta Y^{\bar{a}}} \frac{\delta^l \hat{S}}{\delta Y^*_{\bar{a}}} \eeq
so that the complete path integral, obtained by Hamiltonian methods and by
integrating a part of the extended phase space associated with the
shift symmetry, is given by
\beq \int [d Y^{\bar{a}}] [d \pi_{\lambda a}]
[d Y^*_{\bar{a}}] [d c^{\bar{a}}]
[d {\cal P}^a] [d \bar{{\cal P}}_a ] [d \pi_{\alpha} ]
  \exp\left[\frac{i}{\hbar} Y^*_{\bar{a}} c^{\bar{a}} \right] \hat{U}
\left(  \exp \left[ \frac{i}{\hbar} \hat{S} \right]  \right)\eeq
where the operator $\hat{U}$ is defined by
\beq \hat{U} = \exp \left[
     \frac{\delta^r \Psi_0}{\delta Y^{\bar{a}}} \frac{\delta^l }
     {\delta Y_{\bar{a}}^*}  \right]
\eeq
We now define ${\cal S}$ by
\beq e^{\frac{i}{\hbar} {\cal S}(Y^{\bar{a}},Y^*_{\bar{a}})} = \int
[d {\cal P}^a] [d \bar{{\cal P}}_a ] [d \pi_{\alpha} ]  e^{\frac{i}
     {\hbar} \hat{S}} \label{SBV} \eeq
Since the operator $\hat{U}$ commutes with the integrations done to
define ${\cal S}$, we can rewrite the partition function as
\beq \int [d Y^{\bar{a}}] [d \pi_{\lambda a}]
[d Y^*_{\bar{a}}] [d c^{\bar{a}}]
 \exp \left[ \frac{i}{\hbar} \left( Y^*_{\bar{a}} c^{\bar{a}} + {\cal S}
( Y^{\bar{a}} , Y^*_{\bar{a}} + \frac{\delta^r \Psi_0}{\delta Y^{\bar{a}}
})\right) \right] \eeq
where we used the well-known relation $ \exp \left[ a(y) \frac{\delta}
{\delta x} \right] f(x) = f(x + a(y)) $.
Integrating over the ghost $c^{\bar{a}}$ leads to a
$\delta(Y^*_{\bar{a}})$, removing the anti-fields. We thus recover the
gauge fixing procedure of \cite{BV}, which is to start from an extended
action, function of fields and anti-fields, and to replace the latter
by a derivative of some gauge fermion w.r.t. its conjugate field.

In contrast to \cite{GriGriT}, we do not integrate over the Lagrange
multipliers, but include them together with their momenta in the set of
degrees of freedom of the obtained Lagrangian system. This is done because
they are usually needed to get a covariant Lagrangian in the end, as the
example
of Yang-Mills theories teaches us. There the Lagrange multipliers are
nothing
but the fourth component of the Lorentz-vector $A_{\mu}^a$.

It is now trivial to derive the
  Lagrangian quantum master equation, satisfied by ${\cal S}$. We use the
Fradkin-Vilkovisky theorem, which states that we are allowed to change
$\Psi_0$ to $\Psi_0 + \Delta \Psi$ without changing the partition function
. For
an infinitesimal change in the gauge fermion we thus get
\beq \int [d Y^{\bar{a}}] [d Y^*_{\bar{a}}]
[d c^{\bar{a}}] [d \pi_{\lambda a}]
 e^{+\frac{i}{\hbar} Y^*_{\bar{a}} c^{\bar{a}}}
\frac{\delta^r \Delta \Psi}{\delta Y^{\bar{a}}}  \frac{\delta
  ^l }{\delta Y^*_{\bar{a}}} e^{\frac{i}{\hbar} \hat{{\cal S}}}= 0 \eeq
for any choice for $\Delta \Psi$.
We denoted ${\cal S} (Y^{\bar{a}} , Y^*_{\bar{a}} + \frac{\delta^r
\Psi_0 }{\delta Y^{\bar{a}}}) = \hat{{\cal S}} $. Because of the
arbitrariness of $\Delta \Psi$, this leads after a partial integration
to the well-known Quantum Master Equation \cite{BV}
\beq 0 = (-1)^{\bar{a}} \frac{\delta^l}{\delta Y^{\bar{a}}}
 \frac{\delta^l}{\delta Y^*_{\bar{a}}}
 e^ { \frac{i}{\hbar} \hat{\cal{S}}} = \Delta e^{\frac{i}{\hbar} \hat{\cal
{S}}} \eeq
which is equivalent to the more familiar form
\beq    ( \hat{{\cal S}} , \hat{{\cal S}} ) = 2 i \hbar \Delta \hat{{\cal
S}} \eeq
The so-called anti-bracket is defined as
\beq ( F , G ) = \frac{\delta^r F}{\delta Y^{\bar{a}}} \frac{\delta^l
 G}{\delta Y^*_{\bar{a}}} - \frac{\delta^r F}{\delta Y^*_{\bar{a}}}
\frac{\delta^l G}{\delta Y^{\bar{a}}} \eeq
Instead of gauge fixing the original symmetries as done in (
\ref{gf}), we could also just have added
\beq [ \Psi_0(Y^{\bar{a}}, \pi_{\lambda_a})
+\wt{\bar{{\cal P}}_a \lambda^a} ,
 \Omega ]
 = - \frac{\delta^r \Psi_0}{\delta Y^{\bar{a}}} c^{\bar{a}} + [ \bar{{\cal
 P}}_a \lambda^a , \Omega_0 ] \eeq
where we have done the redefinition (\ref{redef}), and used that
integrating over the collective field and its momentum makes the
tildes disappear.
If we now integrate over the $c^{\bar{a}}$ ghost, we get as gauge
fixing condition \beq \delta(Y^*_{\bar{a}}
 -  \frac{\delta^r \Psi_0}{\delta
Y^{\bar{a}}})\eeq
Redefining \beq Y^*_{\bar{a}} -\frac{\delta^r \Psi_0}{\delta Y^{\bar{a}}}
\rightarrow Y^*_{\bar{a}} \eeq one easily sees that this is exactly the
same
partition function as before.

Let us now try to get a better understanding of ${\cal S}( Y , Y^*)$
defined in
(\ref{SBV}). Following the usual procedure for solving the quantum
master equation \cite{BV}, we expand it in a series in $\hbar$
\beq {\cal S} = S_0 + \sum_{i=1}^{+\infty} \hbar^i M_i \eeq
We will only discuss $S_0$, which satisfies the classical master equation
$(S_0 , S_0) = 0 $. It can be calculated by applying the saddle point
approximation to (\ref{SBV}).
Solving the field equations \beq\frac{\delta \hat{S}}{\delta {\cal P}^a}=
\frac{\delta \hat{S}}{\delta \bar{{\cal P}}_a} = \frac{\delta \hat{S}}
{\delta \pi_\alpha} = 0 \eeq gives us functions
$ {\cal P}^a (Y^{\bar{a}} , Y^*_{\bar{a}} , \pi_{\lambda a})$ and
$ \bar{{\cal P}}_a (Y^{\bar{a}}, Y^*_{\bar{a}} , \pi_{\lambda a}) $ and
$\pi_{\alpha} (Y^{\bar{a}} , Y^*_{\bar{a}} , \pi_{\lambda a})$. Then we
have
\beq S_0 (Y , Y^*, \pi_{\lambda a} ) = \hat{S} \vert_{\Sigma} \eeq where
the
notation $\vert_{\Sigma}$ means that we plugged in the solutions of the
extremum equations. If we now define
\beq ( Y^{\bar{a}} , S_0) \vert_{Y^*=0} = {\cal R}^{\bar{a}}_a c^a \eeq
 then it is trivial to show that
\beq  {\cal R}^{\bar{a}}_a c^a =  [ Y^{\bar{a}}, \Omega_0 ]
 \vert_{\Sigma,Y^*
=0}  \eeq so that indeed the part of $S_0$ linear in the anti-fields
acts as a source term for the BRST-transformations.

This finishes our proof of the equivalence of the Hamiltonian BFV
and the Lagrangian BV formalism. Our guiding principle was that the
Schwinger-Dyson shift symmetry allows for a natural introduction
of anti-fields \cite{us}. Using the prescriptions of the BFV scheme to
implement the Schwinger-Dyson BRST symmetry, we see that the presence of
the
anti-fields need not be restricted to Lagrangian BV. However, the most
obvious
manipulations lead straightforward to an interpretation like that of the
well-known Lagrangian scheme of BV.
The Lagrangian action we get in the end, trivially satisfies the BV
quantum master equation, as a result of the Fradkin-Vilkovisky theorem.
We thus have linked the two principles which assure that the Hamiltonian
and
Lagrangian method can be used for quantising gauge theories at all, namely
,
that the partition functions constructed following their prescription, is
independent of the chosen gauge fixing.

It has been known for some time, that anomalies occur in the BV scheme
when one can not construct a local $M_1$
when trying to solve the master equation at ${\cal O}(\hbar)$\cite
{Troost}. To our knowledge, there have been no studies on how anomalies
would arise in the regularised BFV path integrals. It would be
interesting to see how the momentum integrations play a role in
relating both schemes on that account.

Although both our starting point, the Hamiltonian BFV, and our end
point, the Lagrangian BV, are capable of handling theories with open
algebras, it is not clear how the collective field formalism should
be modified in that case \cite{us}. Clearly, this is one of the major
points to be clarified in the future.

\subsection*{Acknowledgment} It is a real pleasure to thank
P.H. Damgaard for many vivid discussions, and for making my stay at CERN
both enjoyable and mentally profitable. I also want to thank W. Troost
and A. Van Proeyen for sharing their insights on BV.

\end{document}